# Abnormal traffic detection system in SDN based on deep learning hybrid models

Kun Wang[a,b], Yu Fu[a,*], Xueyuan Duan[c,d], Taotao Liu[a], Jianqiao Xu[a]

[a] *Department of Information Security, Naval University of Engineering, Wuhan, 430033, Hubei, China*

[b] *School of Mathematics and Information Engineering, Xinyang Vocational and Technical College, Xinyang, 464000, Henan, China*

[c] *College of Computer and Information Technology, Xinyang Normal University, Xinyang,464000, Henan, China*

[d] *Henan Key Laboratory of Analysis and Applications of Education Big Data, Xinyang Normal University, Xinyang 464000, China*

**Abstract:**
Software defined network (SDN) provides technical support for network construction in smart cities, However, the openness of SDN is also prone to more network attacks. Traditional abnormal traffic detection methods have complex algorithms and find it difficult to detect abnormalities in the network promptly, which cannot meet the demand for abnormal detection in the SDN environment. Therefore, we propose an abnormal traffic detection system based on deep learning hybrid model. The system adopts a hierarchical detection technique, which first achieves rough detection of abnormal traffic based on port information. Then it uses wavelet transform and deep learning techniques for fine detection of all traffic data flowing through suspicious switches. The experimental results show that the proposed detection method based on port information can quickly complete the approximate localization of the source of abnormal traffic. the accuracy, precision, and recall of the fine detection are significantly improved compared with the traditional method of abnormal traffic detection in SDN.

*This document is the results of the research project funded by:
National Key Research and Development Program of China (No.2018YFB0804104) ;
Foundation Strengthening Plan Technical Field Fund Project of the Military Commission Science and Technology Committee(No.2021-JCJQ-JJ-0990).
*Corresponding author
 Email addresses: queen@xyvtc.edu.cn(Kun Wang), fuyu0219@163.com (Yu Fu*)

## 1 Introduction

Smart city integrates the Internet, cloud computing, big data, and artificial intelligence (AI) technologies to break the embarrassing situation of information silos and data segmentation. The Internet is the most important infrastructure for smart cities, and it needs to provide secure, stable, and reliable connection and forwarding functions to meet the rich and diverse network needs of smart cities. Software defined network (SDN) through the unified scheduling of resources and automatic distribution of services, to solve the traditional network computing and storage resource utilization rate is low, high operation and maintenance costs, so SDN is increasingly favored by network operators and equipment manufacturers.

SDN separates the traditional data plane and forwarding plane decouples the forwarding and control of network traffic data, and enables the network to develop independently of hardware devices[1]. The SDN architecture can be divided into the control plane, the forwarding plane, and the application plane, in which the controller in the control plane is the decision-making core of the SDN, which is responsible for the development of the network forwarding policy, the scheduling of the communication unit, and the optimal allocation of network resources. the forwarding plane is mainly composed of network forwarding devices such as switches and routers, which mainly perform traffic forwarding operations according to the forwarding rules issued by the control plane. the application plane is mainly composed of network applications, such as traffic control, responsible for balancing, and other services[2]. The various planes of SDN are connected through common interface protocols, and the systematic network architecture and good perception and control capability of the network make SDN increasingly favored by network operators and equipment vendors.

However, the open design of SDN also provides many facilities for network attackers. Take the most widely used OpenFlow protocol SDN as an example, after the switch receives a new flow packet, it first compares it with its flow table, and then performs according to the rules after it finds the corresponding flow table item. if it does not find the corresponding flow table item, the switch encapsulates the packet's header information in PackIn message and sends it to the controller to ask for a disposal method. The controller parses the PackIn message, formulates forwarding rules and sends them to the switch. the switch generates flow table entries according to the rules and completes the disposal of the packet[3]. This process does not determine the nature of the packet, i.e., the OpenFlow switch accepts arbitrary forwarding requests, then the attacker simply sends a large number of packets containing random header information to the OpenFlow switch in a short period, which will lead to the switch flow table overload, cache overflow, controller resource exhaustion, link congestion, etc., which will affect the quality of service of the SDN, and in severe cases, may cause the whole network to be paralyzed[4]. Therefore, the security problem of SDN itself is unavoidable and urgent in the development of SDN and has become a hotspot in the field of SDN research gradually.

Abnormal traffic detection can effectively detect the abnormal data of traffic, so as to discover the possible attacks in the network, which is the basic means of network security protection. SDN has a different network architecture and workflow from traditional networks, and the organization of network attacks is also significantly different, so the network abnormal traffic detection methods in the SDN environment also have special characteristics[5]. However, most of the current research on SDN abnormal traffic detection[6-9], which is based on traditional abnormal traffic detection methods, generally suffers from the problems of complex algorithms, modeling difficulties, time-consuming detection, etc., and these research efforts do not well reflect the network characteristics and detection requirements of SDN.

In this paper, by systematically analyzing the characteristics of SDN structure and exploring the characteristics of traffic changes caused by network attacks, a hierarchical detection system for abnormal traffic in SDN based on deep learning is proposed, which firstly achieves rough-grained detection of anomalies based on the changes in the statistical characteristics of the traffic on the switch ports. Then the traffic data on the switch ports are collected and the traffic features are extracted with the CIC-FlowMeter tool, and then classifies the feature data using a multi-frequency classifier based on wavelet transform and neural network to achieve fine-grained detection of abnormal flows. The system can quickly complete the initial detection of abnormal traffic with less computational expenditure and without generating additional communication data. when an abnormality is detected, the fine-grained detection method based on multi-frequency features is launched, which can ensure the accurate detection of abnormal data, and this hierarchical detection design may be a better balance between the need for swiftness and accuracy in abnormal detection.

In summary, the major contributions of this paper are the following:
- Using wavelet decomposition technique to process the feature data of suspicious traffic, and combining convolutional neural network and recurrent neural network to construct the abnormal traffic detection model.
- Compared with other similar detection

methods, our proposed MFDLC detection method uses fewer data features and achieves higher detection accuracy.
- Our detection method is tested on different network datasets, and the results show that our method achieves excellent detection results.
- Our method also has the potential to accomplish real-time detection due to the implementation of a two-stage abnormal detection approach.

The rest of this paper is organized as follows. Section II shows the related work. Section III presents a detailed description of the proposed hierarchical abnormal detection system. Section IV describes the experimental environment and experimental data. Section V analyses the detection performance of the model. and finally, Section VI concludes the paper and presents future research work.

## 2 Related Work

### 2.1 Attacks and Traffic Characteristics in SDN

The forwarding plane and control plane of traditional networks are tightly coupled in the same device, and administrators are unable to directly manipulate forwarding behaviors. SDN separates the control plane from the forwarding plane, and open operational semantics facilitate device access to the network. Administrators can use self-designed software to plan arbitrary forwarding behaviors in the network through programmable interfaces, thus achieving the definition of the entire network, the so-called Software Defined Network. The basic architecture is shown in Fig. 1.

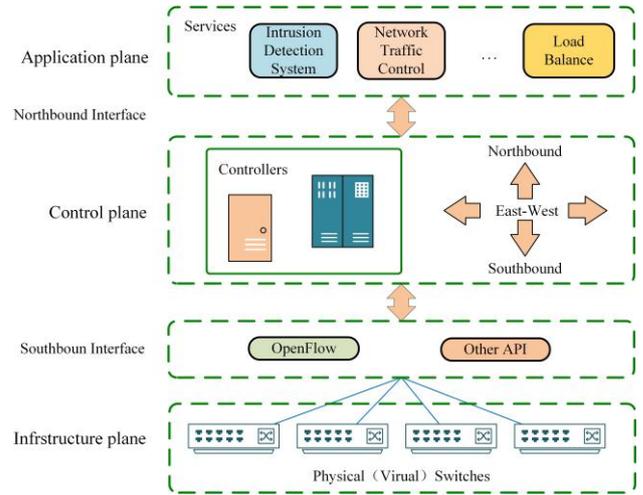

**Fig. 1.** SDN architecture

As can be seen, SDN is a 3-layer plane and 4-way interface organization, in which the south interface is responsible for the communication between the forwarding plane and the control plane, and the OpenFlow protocol is the most widely used south interface. the east-west interface can achieve the information interaction between the controllers, while the north interface has not been standardized yet [10]. Due to the open design of SDN, the control plane and forwarding plane are more susceptible to network attacks, the most common and easy to organize is the DoS (DDoS) attack, the specific attack process and possible effects are shown in Fig. 2.

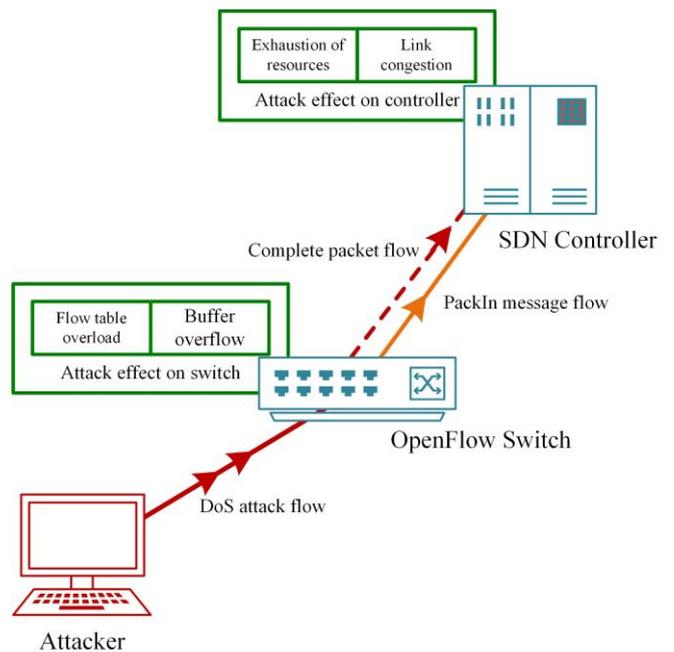

**Fig. 2.** DoS attack process and effect against SDN

DoS attacks against the forwarding plane can trigger flow table overload and cache overflow problems in switches. Since a switch's main role in the network is to provide forwarding services and its own storage space is very limited, when the flow table space is filled with a large number of meaningless flow table entries, it will not be able to provide forwarding services for new normal traffic. In addition, the data parsing capability of the controller is limited, and when a large number of PackIn messages are sent from the switch, its own computing and storage resources will be consumed very quickly, and it will no longer be able to handle legitimate forwarding requests [11]. The communication link between the controller and the switch may also become a bottleneck in the network. Usually, when the switch receives a new packet, it sends only the header information encapsulated in the PackIn message to the controller, and the packet's load will be temporarily stored in its own cache. When the switch cache is filled and overflow occurs, the load data will also be packaged and sent to the controller, and a large amount of concurrent uplink data may also trigger congestion on the communication link. These attacks are usually accompanied by changes in traffic data in the communication link, as shown in Table 1.

**Table 1** Attack types and traffic characteristics against SDN

| Aggression efficacy | Organization mode | Traffic characteristics |
| --- | --- | --- |
| Flow table overload [12] | Attackers use DoS attacks or DDoS attacks to generate a large number of mismatched flows, and the controller generates a lot of new forwarding rules for the switch, which leads to the exhaustion of the storage space of the switch flow table and can no longer provide normal forwarding services for new legitimate flows. | A lot of mismatched flows flood into the switch in a short time, and the switch forwards a lot of Packet_In messages to the controller in a centralized way. |
| buffer overflow [13] | Initiated by a DoS attack or DDoS attack program, when the switch receives large numbers of mismatched flows, it will temporarily store the load of all data packets in the cache. If the switch cache is full, packet loss will occur, resulting in the loss of normal flow information. | In a short time, large numbers of streams are transmitted to the switch, but the traffic data that the switch can forward normally is relatively small. |
| exhaustion of resource [14] | Attackers independently generate or control puppet machines to generate some meaningless flows to trigger the switch sending large numbers of Packet_In messages to the controller which makes the controller overloaded and incapable of responding to normal service requests. | Large numbers of Packet_In packets are transmitted in the link from the switch to the controller. |
| link congestion [15] | Attackers send large numbers of meaningless streams or TCP messages to the controller, which greatly consumes the link bandwidth from the forwarding plane to the control plane, making normal service requests unable to reach the controller. | In short time, the link between the switch and the controller transmits high-speed upstream traffic. |

SDN may also suffer from network attacks such as switch hijacking, ARP attacks, viruses, and worms. However, DoS attacks are diverse and easy to organize. Attackers can easily exhaust the available resources of SDN by manipulating the puppet machine to send large numbers of seemingly normal packets such as TCP, UDP, ICMP, or DNS to the switch. It can be said that DoS attacks are the most common and direct network attacks that SDNs face, so detecting DoS attacks is also the main task of maintaining SDN security.

## 2.2 Abnormal traffic detection in SDN based on deep learning

Deep learning is a kind of machine learning based on a neural network algorithm, which can learn the high-level characteristics of data from unstructured data layer by layer by using a multi-layer neural network, thus having strong representation ability. Like machine learning, deep learning can be divided into supervised learning, unsupervised learning, and semi-supervised learning according to whether the samples are labeled or not. At first, neural networks were mostly fully connected structures, also known as a multi-layer perceptron, and then gradually evolved into various structural modes. Therefore, according to the used neural network model structure, SDN abnormal traffic detection can be divided into three types: based on convolutional neural network (CNN)[16], recurrent neural network (RNN)[17], autoencoder network (AE)[18], deep belief network (DBN)[19], and generative adversarial network (GAN)[20], as shown in Fig. 3.

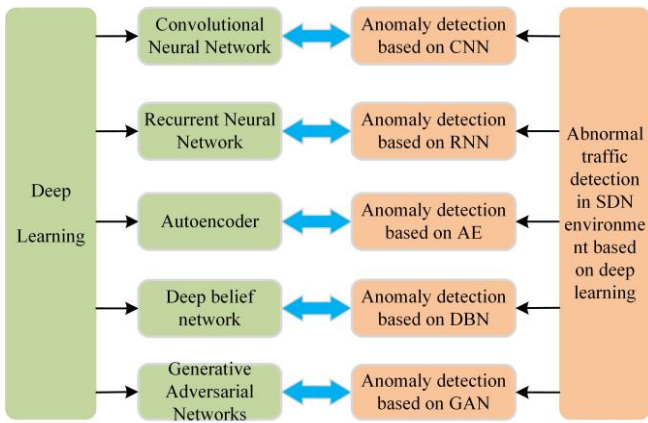

**Fig. 3.** Abnormal traffic detection method in SDN environment based on deep learning

CNN and RNN are both typical supervised learning models, and they play a role in abnormal traffic detection by virtue of their excellent spatial feature extraction ability and temporal feature extraction ability respectively. For example, Literature [21] proposes a network traffic abnormal detection model based on regularized CNN and LSTM. Firstly, 48 features of each network flow in InSDN, a special data set for SDN traffic data analysis, are converted into a two-dimensional image format of $8\times6$, and the spatial characteristics of network flows are learned by CNN, and then the temporal characteristics of network flows are learned by LSTM network, thus accurately depicting the spatial and temporal characteristics of network traffic behavior. This method is better than using the CNN or LSTM model alone to detect abnormal traffic. However, because the original sample category is unbalanced, it is necessary to balance the training data manually, which leads to the sub-optimal solution of the supervised learning detection model [22]. In addition, it is difficult to obtain large-scale clearly available tagged data in reality, so the abnormal traffic detection technology with supervised learning is not as popular as the unsupervised learning or semi-supervised learning method.

DBN, AE network, and GAN are typical unsupervised learning models, all of which have self-coding functions and can automatically label unlabeled data samples [23], so they are widely used in the field of abnormal traffic detection. Wang et al. [24] proposed a method to detect DDoS attacks in SDH based on DBN. Firstly, the characteristics of OpenFlow switch flow table entries were extracted, and then the trained DBN was used to detect whether DDoS attacks existed in the network. Shone et al. [25] proposed an abnormal traffic detection model based on asymmetric depth self-encoder (NDAE) with unsupervised feature learning. The high-order features of traffic data were extracted layer by layer by a multi-hidden layer self-encoder, and finally, the abnormal traffic was automatically identified by the classifier. However, due to the prior assumption of data distribution, the accuracy of an unsupervised learning algorithm for abnormal traffic detection is usually low, so people begin to study semi-supervised learning technology which uses a limited number of labeled samples and large numbers of unlabeled samples to train together.

Abnormal traffic detection of semi-supervised learning is a combination of supervised learning and unsupervised learning methods, which can usually be divided into two situations: one is to use a small amount of labeled data and some unlabeled data to model together. Wang et al. [26] use a small amount of labeled data and a large amount of unlabeled data for model training. First, they truncate the data of SDN traffic packets or fill them with "0" to convert them into data packet byte vectors with the same length. Then multiple standardized data packet vector samples are input into the semi-supervised model, and the generator and discriminator are alternately trained. This multi-sample simultaneous training method can effectively overcome the problem that the number of sample labels is insufficient, and it is difficult to model. The other is to fully train the depth self-coding model with normal sample data [27], and the trained model can reconstruct the normal samples well, so the samples with large reconstruction errors can be decided as abnormal.

In addition, the hybrid abnormal traffic detection algorithm [28-30], which integrates various model structures, can obviously improve the detection ability of abnormal traffic by means of the strategy of task division or comprehensive integration of detection results and has gradually become a research hotspot in the field of network security. Table 2 shows the performance and characteristics of some abnormal traffic detection methods in an SDN environment based on deep learning.

**Table 2** Abnormal traffic detection method in SDN based on deep learning

| Algorithm name | Algorithm description | Data set | Accuracy | Advantage | Disadvantages |
| --- | --- | --- | --- | --- | --- |
| CNN-SoftMax [16] | CNN is introduced into the regularization method of the standard deviation of the weight matrix to improve generalization ability. | InSDN | 0.982 7 | Higher detection accuracy, using GPU to accelerate training. | Long training time, and feature specification may cause the loss of independent variables. |
| DNN-LSTM [17] | Deep LSTM learns the time correlation of traffic data, and the classifier completes abnormal identification. | CIC-IDS2017 | 0.993 2 | High accuracy and can detect the traffic generated by multiple types of malwares. | Complex algorithm, data in an unreal SDN environment. |
| SSAE-SVM [18] | Information entropy can quickly detect suspicious traffic and stacking sparse AE and support vector machine for confirmation. | Self-collected data set | >0.98 | Detection data conforms to the characteristics of the SDN environment, and the detection accuracy is high. | Complex model, large amount of calculation, and high network cost. |
| DBN-TWD [19] | DBN extracts the characteristics of traffic and classifies it, and the K-nearest neighbor method reclassifies the boundary data. | Self-collected data set | 0.967 2 | K-nearest neighbor algorithm is used for reclassification to improve the accuracy. | Complex model design, high computational consumption. |
| GAN [20] | Continuous monitoring of IP flows to detect DDoS attacks, confrontational training to reduce the sensitivity of the detection system to attack disturbances. | CIC-DDoS2019 | 0.943 8 | Effective detection of common attacks, with certain anti-interference ability. | Data in unreal SDN environment, the system adds extra load and overhead. |
| CNN-LSTM [21] | Using CNN and LSTM to extract the spatial and temporal features of data respectively. SoftMax classifier completes | InSDN | 0.963 2 | Accuracy is higher than that of the CNN or LSTM model alone. | The algorithm is relatively complex, with small throughput and large delay, so it is difficult to meet the |

| Algorithm name | Algorithm description | Data set | Accuracy | Advantage | Disadvantages |
|---|---|---|---|---|---|
| | the detection of abnormal traffic. | | | | requirements of online detection. |
| DBN [24] | Extract the characteristics of OpenFlow switch flow table entries, and DBN detects DDoS attacks. | Self-collected data set | 0.950 0 | Compared with the traditional method, the detection accuracy is improved. | More detection time spending |
| NDAE [25] | Multi-hidden layer self-encoder extracts high-order features of traffic data layer by layer and uses classifier to achieve abnormal traffic detection. | UNSW-NB15 | 0.891 2 | Detection data is easy to obtain, and high-order features of the data can be extracted. | Data in unreal SDN environment has complex model structure and low detection accuracy. |
| ByteSGAN[26] | Using standardized data to train the model alternately to improve the detection accuracy and generalization ability. | ISCX-IDS2012 | 0.991 8 | Using the generated samples for effective training, the encrypted traffic can be detected. | Data in unreal SDN environment and the algorithm is complex and has no advantage when the samples are sufficient. |
| GRU-DAE [27] | GRU and DAE combine to use supervised abnormal detection with selective mechanism to assist semi-supervised abnormal detection. | NSL-KDD | 0.902 1 | Can detect unknown network attacks | Data cannot reflect the characteristics of SDN environment, and the detection accuracy is low. |
| 1D-CNN&2D-CNN [28] | Integrate CNN, LSTM, GRU and other deep learning models to build a multi-model abnormal traffic detection framework. | CIC-IDS2017 | 0.997 7 | Integrating the feature extraction advantages of multiple depth models improves the classification effect of SDN traffic. | Data cannot reflect the characteristics of SDN environment, and the algorithm is complex, and the detection takes a long time. |
| CU-DNNGRU& Cu-BLSTM [29] | Using DNNGRU and BLSTM hybrid framework to detect complex threats and malware on the Internet of Things environment and using GPU to improve computing performance. | CSE-CIC-IDS2018 | 0.998 7 | High detection accuracy and high speed. | Data cannot reflect the characteristics of SDN environment, and the model is huge and has many parameters. |
| BDLSTM-AE [30] | Combining bidirectional long-term and short-term memory networks with self-coding networks, a deep stack self-coding abnormal detection model is constructed. | ISCX-IDS2012 UNSW-NB15 | 0.993 0 0.999 4 | Model has strong stability and high detection accuracy. | Data cannot reflect the characteristics of SDN environment, and the model is complex, and the detection takes a long time. |

As can be seen from Table 3, the current research results of SDN abnormal traffic detection based on deep learning have the characteristics of diverse model structures and high detection accuracy. However, each detection method also has its own shortcomings and deficiencies, which are as follows: First, most detection systems are deployed in the server where the controller is located, which will occupy the resources of the controller during detection, leading to the decline of the controller's performance or even its failure to work normally. Second, the detection algorithm is complex, there are many operational data, the calculation cost is large, and the detection time is long, so it is difficult to find the abnormal traffic in the network in time. Third, in addition to self-collected data sets, most of them use traditional network traffic data sets, and the data cannot reflect the characteristics of SDN environment, so it is difficult to evaluate the detection effect of detection methods on real SDN.

In addition, the current abnormal traffic detection models based on deep learning are mostly single architecture, which does not fully mine the rich potential characteristic information of network traffic. Signal experts found that the time-varying signals of normal traffic and abnormal traffic are quite different in frequency [31], Some scholars tried to design an abnormal detection algorithm by using the time-frequency domain characteristics of traffic. Cheng et al. [32] proposed a multi-scale LSTM model to capture abnormal behavior in border gateway protocol traffic. They first used discrete wavelet transform to obtain traffic representations on multiple scales, and then used LSTM to learn the time dependence in these representations. Duan et al. [33] used sliding window and wavelet transform technology to obtain multiple feature subsequences at different time and decomposition scales from the original traffic sequence and input them into a classifier composed of stacked AE and CNN residual network to detect abnormal traffic samples.

## 3 Hierarchical abnormal traffic detection system

### 3.1 System overview

The hierarchical abnormal traffic detection system we designed is deployed on the server where the SDN controller is located. It consists of an information collection module, a hierarchical abnormal traffic detection module, and an abnormal traffic disposal module. The structural framework is shown in Fig. 4.

The traffic information acquisition and analysis module has two working modes: port characteristic acquisition and flow characteristic acquisition. Among them, port characteristic acquisition mainly completes the statistical characteristic information of the packets or traffic flowing through each OpenFlow switch port in general; traffic characteristic acquisition is launched when abnormalities are initially detected in the network, and mainly uses packet capturing tools to capture all the traffic data flowing through the suspicious switch; and then uses the traffic analysis tool to extract the traffic characteristic information. The hierarchical detection model is also divided into two stages: one is to complete the initial detection of abnormal traffic data in the network using port feature information; if an abnormality is detected in a switch, the next stage of deep detection is initiated, and the extracted traffic features of the switch are accurately detected using the deep learning model. The abnormal traffic disposal module mainly completes the traceability and location of abnormal traffic, and prevents abnormal hosts from sending attack flows into the SDN from the source by prohibiting the switch from sending PackIn messages and dropping the received new flows into the default ports.

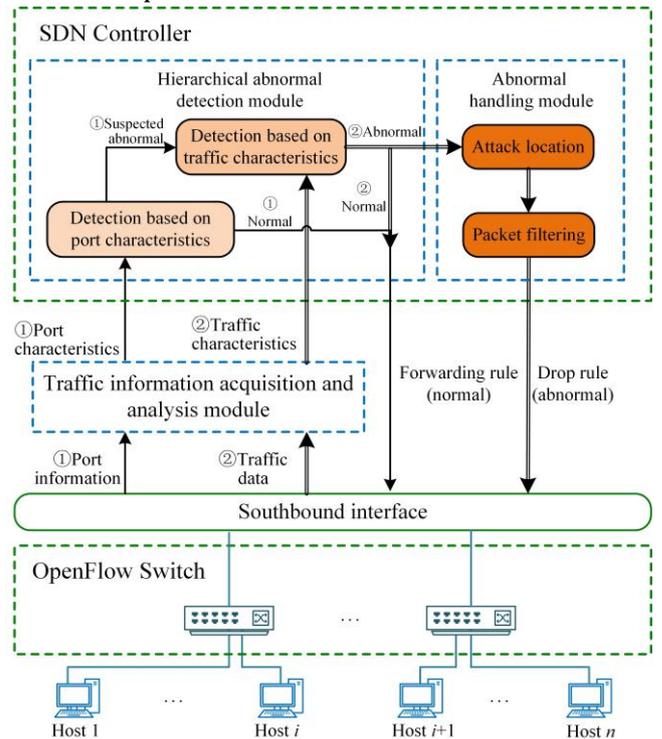

**Fig. 4.** Hierarchical framework of abnormal traffic detection system

### 3.2 Hierarchical abnormal traffic detection module

The hierarchical detection module uses two detection functions to continuously monitor the traffic data flowing through the OpenFlow switch. Firstly, the abnormal traffic is preliminarily detected according to the information of the switch port. When the suspicious situation of the switch is detected, the collected traffic characteristics are fine detected by using the multi-frequency depth model. Before introduction in detail, the related symbols used in this paper are explained, as shown in Table 3.

**Table 3** List of Common Symbols

| symbol | meaning |
| --- | --- |
| $Num_{FlowIn}$ | Number of network flows flowing into the switch |
| $Num_{FlowOut}$ | Number of network flows flowing out of the switch. |
| $Num_{PacketIn}$ | Number of PacketIn packets forwarded from switch to controller |
| $Rate_{PacketIn}$ | Ratio of the number of network flows flowing into the switch to the number of PacketIn packets forwarded from the switch to the controller (inflow-forwarding ratio) |
| $Rate_{FlowIO}$ | Ratio of network flow out of the switch to network flow into the switch (normal forwarding ratio) |
| $\varphi_{CNN}^{i}$ | ith -level CNN model function |
| $\theta_{CNN}^{i}$ | ith -level CNN model parameter |
| $\phi_{LSTM}^{i}$ | ith -level LSTM model function |
| $\theta_{LSTM}^{i}$ | ith -level LSTM model parameter |

### 3.2.1 Abnormal detection based on port information

Periodically, the ports of the switch count the number of network flows and the number of packets flowing into or out of the switch, as well as the number of PacketIn messages forwarded by the switch to the controller. Under normal circumstances, flows with corresponding matching rules in the switch flow table can be processed normally, and flows without matching rules need to ask the controller for processing methods, but the number of such flows is generally not very large, especially for SDN switches running for some time, which $Num_{FlowIn} \gg Num_{PacketIn}$, the ratio between them is a bigger value.

However, when a DDoS attack is launched, a large number of new flows arrive at the switch requesting forwarding operations within a short period of time. Because the source IP or destination IP of these flows is randomly generated, there is no corresponding flow table entry in the switch flow table, so the switch will send lots of PacketIn messages to the controller to ask how to deal with them. Every incoming new flow will generate a PacketIn message sent to the controller, so the ratio of the number of network flows flowing into the switch to the number of PacketIn packets forwarded from the switch to the controller, that is, the inflow-forwarding ratio, will be much lower than normal, and the inflow-forwarding ratio can be expressed as

$$Rate_{PacketIn} = \frac{Num_{FlowIn}}{Num_{PacketIn}} \qquad (1)$$

In addition, even if there is a DDoS attack in the network, some normal flows with matching rules in the switch can still be forwarded correctly before the switch is completely blocked, but their proportion (normal forwarding ratio) in all network flows flowing into the switch will be greatly reduced compared with the normal situation, that is, the normal forwarding ratio

$$Rate_{FlowIO} = \frac{Num_{FlowOut}}{Num_{FlowIn}} \qquad (2)$$

When both $Rate_{PacketIn}$ and $Rate_{FlowIO}$ of a switch are less than the determining threshold, it can be determined that there is abnormal traffic caused by network attack in the switch.

### 3.2.2 Abnormal detection based on traffic data

The detection of this phase relies on the MFDLC (multi-frequency deep learning classifier), which is started when an anomaly is detected in a switch based on port information. Firstly, Wireshark, a traffic collection program, is used to capture all the traffic data flowing through the switch, and then CIC-FlowMeter analysis tool is used to extract the characteristics of the traffic; then after data preprocessing, it is sent to MFC to achieve the accurate detection of abnormal traffic. MFC is a combination of wavelet transform technology and deep learning traffic classification model, which can extract the time-frequency features of the traffic data using wavelet transform, extract the spatial features of the data using CNN and extract the temporal features of the data using LSTM, and comprehensively analyze the traffic data from the three dimensions of "frequency domain, space, and time", and finally complete the classification of the data type by the SoftMax classification function, the structure of which is shown in Fig. 5.

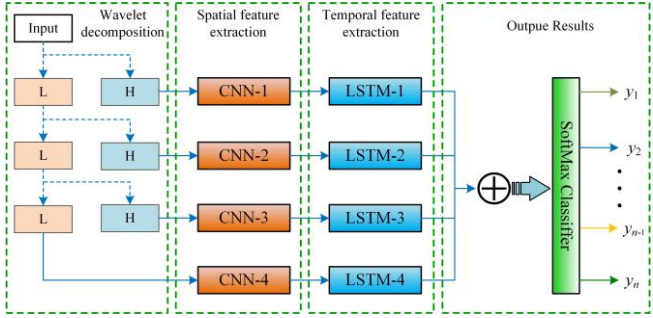

**Fig. 5.** MFC architecture

We know that the time correlation of hidden points in sequence data is closely related to frequency. Large-time-scale correlations, such as long-term trends, are usually located at low frequencies, while small-time-scale correlations, such as short-term disturbances and events, are usually located at high frequencies. In order to deeply explore the correlation in the flow sequence, we apply wavelet decomposition to the input sequence $\mathrm{X} = \{x_1, x_2, ..., x_k\}$, and get the low-frequency components and high-frequency components of the *ith-level* as follows

$$x^l(i) = \{x_1^l(i), x_2^l(i), ..., x_k^l(i)\} \quad (3)$$

$$x^h(i) = \{x_1^h(i), x_2^h(i), ..., x_k^h(i)\} \quad (4)$$

Since we use the decomposed sequence of the original sequence without wavelet reconstruction, we do not need to use downsampling practices when decomposing again. After n decompositions, n+1 subsequences with the same dimensions as the original sequence can be finally obtained, and the set of sequences can be represented as

$$\chi(n) = \{x^h(1), x^h(2), ..., x^h(n), x^l(n)\} \quad (5)$$

Each sub-sequence is converted into a 2D graphical format and input into an independent CNN for spatial characteristic extraction, and the results obtained from each sub-sequence through a series of convolutional operations are then input into an independent LSTM network to further extract the temporal characteristics of each sub-sequence

$$z^i = \phi_{LSTM}^i [\varphi_{CNN}^i(x^i, \theta_{CNN}^i), \theta_{LSTM}^i] \quad (6)$$

In which, $x^i$ is the subsequence decomposed by the *ith-level* wavelet and the input of the *ith-level* CNN, and $z^i$ is the output subsequence transformed by CNN and LSTM. Finally, the results of each subsequence operation are summarized, and the classification of traffic samples is completed by the SoftMax function.

$$y' = \mathrm{Softmax}(\frac{1}{m}\sum_{i=1}^{m} z^i) \quad (7)$$

### 3.3 Abnormal handling module

When attackers launch DoS attacks using flood flows, they usually use randomly generated forged addresses, which makes it difficult to trace back using addresses. Therefore, we utilize the switch port number as the sample's identifier when collecting traffic information. When an abnormal sample is detected, the controller determines the switch where the abnormality occurs based on the sample's identifier, issues a command to the switch to reject new traffic from that port into the SDN, and prohibits that switch from sending PackIn messages for a set period of time; the switch introduces new traffic flowing into that port into the default port for discarding, which is implemented with the following algorithm.

| Algorithm of abnormal handling |
|---|
| Input：Attacking link information (Switches, ports, etc.) |
| Output：Handling rules for new traffic (Ports, handling methods, etc.) |
| 1. IF $Port_A \in \{Port_i\}$  // If the abnormal port is a switch port. |
| 2. NoPktIn=true  //Prohibit the switch from sending the PacketIn message. |
| 3. set_new_dst_port=default   //Import a new traffic into the default port. |
| 4. fp_action=drop   // Discard the new traffic. |

5.　　$t = T_{lim}$　　// $T_{lim}$ is the set limit time.

6.　　RETURN　the handling rule for new traffic.

## 4　Experiment dataset and data preprocessing method

### 4.1　Experiment dataset
#### 4.1.1　Dataset selection

The performance of anomaly detection techniques relies heavily on the quality of the training data. Most of the currently available public datasets, such as ISCX-IDS2012, UNSW-NB15, CIC-IDS2017, and CIC-DDoS2019 are derived from traditional network environments and do not really reflect certain characteristics of network traffic in SDN. For example, the sharing of SDN controllers with network devices increases the chances for attackers to execute various types of attacks on data communication links or on the SDN controllers themselves, which are generally difficult to detect because the attackers are usually connected to the network in a capacity that the SDN considers legitimate. Therefore, to better characterize the experiments in the SDN environment, we use the InSDN dataset [34]. InSDN is traffic data in SDN published by the University of Dublin, Ireland, and contains the most common types of current attacks, mainly DoS attacks, DDoS attacks, Probe, Brute Force Breach, Botnets, Web Attacks, etc. Normal Traffic Data is generated by the popular current application services, such as HTTP, HTTPS, DNS, E-Mail, FTP, SSH, etc. InSDN provides two formats, PCAP and CSV, in which the CSV file is composed of more than 80 statistical features extracted by using the CIC-FlowMeter traffic analysis tool. It consists of three files containing a total of 343,889 sample records, of which 275,465 are attack samples and 68,424 are normal samples, and the specific information is shown in Table 4.

**Table 4** Sample distribution of InSDN dataset

| Sample category | Sample number | Sample ratio (%) |
|---|---|---|
| DDoS | 121 942 | 35.460 |
| DoS | 53 616 | 15.591 |
| Probe | 98 129 | 28.535 |
| Brute-Force-Attack | 1 405 | 0.409 |
| Web-Attack | 192 | 0.056 |
| BotNet | 164 | 0.048 |
| U2R | 17 | 0.005 |
| Normal | 68 424 | 19.897 |
| **Total** | **343 889** | |

#### 4.1.2　Data characteristic selection

Processing raw data is a great advantage of deep learning. However, behind this advantage, powerful high-performance computing power is needed as support, and it will also consume more computing time. In order to reduce the calculation pressure and shorten the calculation time, according to the research results of Krishnan et al. [35], we select 48 features from more than 80 features counted by CIC-FlowMeter as the data of model training and testing experiments, as shown in Table 5. In fact, some of these selected features can be obtained directly from the data layer equipment through the SDN controller, while others can be designed manually from the data obtained by the controller.

**Table 5** Subset of InSDN Dataset Characteristics

| No. | Attribute name | No. | Attribute name |
|---|---|---|---|
| 1 | Protocol | 25 | Fwd IAT Total |
| 2 | Flow duration | 26 | Bwd IAT Min |
| 3 | total Fwd Packet | 27 | Bwd IAT Max |
| 4 | total Bwd packets | 28 | Bwd IAT Mean |
| 5 | total Length of Fwd Packet | 29 | Bwd IAT Std |
| 6 | total Length of Bwd Packet | 30 | Bwd IAT Total |
| 7 | Fwd Packet Length Min | 31 | Fwd Header Length |
| 8 | Fwd Packet Length Max | 32 | Bwd Header Length |
| 9 | Fwd Packet Length Mean | 33 | FWD Packets/s |
| 10 | Fwd Packet Length Std | 34 | Bwd Packets/s |
| 11 | Bwd Packet Length Min | 35 | Packet Length Min |
| 12 | Bwd Packet Length Max | 36 | Packet Length Max |
| 13 | Bwd Packet Length Mean | 37 | Packet Length Mean |
| 14 | Bwd Packet Length Std | 38 | Packet Length Std |
| 15 | Flow Bytes/s | 39 | Packet Length Variance |

| No. | Attribute name | No. | Attribute name |
|---|---|---|---|
| 16 | Flow Packets/s | 40 | Average Packet Size |
| 17 | Flow IAT Mean | 41 | Active Min |
| 18 | Flow IAT Std | 42 | Active Mean |
| 19 | Flow IAT Max | 43 | Active Max |
| 20 | Flow IAT Min | 44 | Active Std |
| 21 | Fwd IAT Min | 45 | Idle Min |
| 22 | Fwd IAT Max | 46 | Idle Mean |
| 23 | Fwd IAT Mean | 47 | Idle Max |
| 24 | Fwd IAT Std | 48 | Idle Std |

### 4.2 Data preprocessing

Normally, there may be some problems in traffic data, such as data defects, format errors, large dimension differences, etc. Therefore, before inputting into the deep learning abnormal traffic detection model, some preprocessing operations are needed, including data cleaning, eigen encoding, data normalization, and so on. Data cleaning mainly eliminates data samples with eigenvalue defects in data sets. If the sample has a high defect rate (greater than 80%) and low importance, it will be deleted directly. However, for the data with a small defect rate and important data, the average interpolation method is usually used to repair the data to ensure the integrity of the data samples. Eigen coding is mainly to encode non-numerical eigenvalues to make them computer-recognizable values. This paper adopts the method of one-hot encoding. The main purpose of numerical normalization is to scale the data according to a certain proportion in order to avoid the adverse impact of inconsistent data dimensions on the detection results. In this study, the Min-Max normalization method is adopted, and the calculation method of eigenvalue can be expressed as follows.

$$x_i^{'} = \frac{x_i - x_{min}}{x_{max} - x_{min}} \quad (8)$$

Where $x_i$ is the eigenvalue before normalization, $x_i^{'}$ is the eigenvalue after normalization, $x_{max}$ and $x_{min}$ are the maximum and minimum values of the eigenvalues before normalization, after normalization, all the eigenvalues are mapped in the interval [0,1], with the maximum eigenvalue being 1 and the minimum eigenvalue being 0.

## 5 Experimental results and analysis

### 5.1 Experimental setup

The experimental environment in this paper is configured as Core i9-12900F, 128GB RAM and NVIDIA RTX3090. CUDA11.2, Pytorch1.8. The network of SDN is built on Ubuntu 16.04LTS operating system, and the experimental environment is built with Mininet simulator and POX controller. The network topology is shown in Fig.4. Mininet simulates four switches and four hosts connected to each switch and communicates with the POX controller. The address information of controllers and hosts is shown in Table 6.

**Table 6** Address information of controller and host

| Name | IP address | Roles |
|---|---|---|
| controller | 12.0.0.1 | Flow control |
| host 1 | 12.0.0.11/15/19 | Attacker |
| host 2-4 | 12.0.0.12-12.0.0.14 | Normal user (HTTP) |
| host 5-8 | 12.0.0.15-12.0.0.18 | Normal user (HTTPS) |
| host 9-12 | 12.0.0.19-12.0.0.22 | Normal user (FTP) |
| host 13-16 | 12.0.0.23-12.0.0.26 | Normal user (SSH、email) |

### 5.2 The evaluation metrics

To comprehensively evaluate the detection efficiency of the proposed method for abnormal traffic in the network, five evaluation indexes such as Accuracy, Precision, Recall, FPR (False Positive Rate), and F1 value are used：

$$Accuracy = \frac{TP + TN}{TP + TN + FP + FN}$$

$$Precision = \frac{TP}{TP + FP}$$

$$Recall = \frac{TP}{TP + FN}$$

$$F1 = \frac{2 \times Precision \times Recall}{Precision + Recall}$$

$$FPR = \frac{FP}{FP + TN}$$

Where $TP$, $TN$, $FP$ and $FN$ are the

interrelationships between the real classification and the predicted classification of the sample, and they can be represented by Table 7 of the relation matrix.

**Table 7** Relationship matrix between real classification and predicted classification

|  | Prediction abnormality | Prediction normal |
|---|---|---|
| True abnormality | TP | FN |
| True normal | FP | TN |

### 5.3 Model parameters setting

The multi-frequency deep learning classifier proposed by us in Section 3.2 is designed by combining wavelet transform with deep learning. The wavelet basis function uses the Daubechies wavelet (DB), and the wavelet decomposition level is 3. The CNN network of deep learning consists of two convolution layers, using a $3\times 3$ convolution kernel, and each convolution layer is followed by a $2\times 2$ maximum pooling layer, Dropout algorithm is used to prevent over-fitting. The number of nodes of LSTM is 128. The mean square error is used as the loss function, the small batch gradient descent method is used to minimize the loss function, and the BP algorithm is used to update the model parameters. Other hyper-parameter settings are shown in Table 8.

**Table 8** MFDLC hyperparameter setting

| Parameter name | Parameter value |
|---|---|
| wavelet basis function | DB4 |
| wavelet decomposition level | 3 |
| CNN- convolution kernel | 3×3 |
| CNN- activation function | Relu |
| CNN- Convolution Layer Output | 32、64 |
| CNN-Dropout | 0.2、0.3 |
| max pooling | 2×2 |
| LSTM- max pooling | 128 |
| LSTM- activation function | sigmoid、tanh |
| regularization | L2 |
| learning rate | 0.01 |

### 5.4 Experimental results and analysis

The experiment is divided into two stages. The first stage mainly verifies the performance of abnormal traffic detection methods based on port information, and the second stage mainly evaluates the performance of a multi-frequency deep learning classifier.

#### 5.4.1 Abnormal detection based on port information

In this stage of the experiment, host 1 is set as the attack source, and the Hping3 tool is used to send SYN pulses containing multiple randomly forged target IP addresses periodically and continuously to the switch with a traffic intensity of 20Mb/s (Send 1s- Idle 5s). Other hosts send normal packets with random intensity according to the configured service. The traffic information acquisition and analysis module continuously count the port information of each switch for a period of 1 second, converts it into the data format required for detection, and inputs it to the abnormal detection module. The experiment lasted for 1 hour, and the attacking host mainly launched 600 attacks. Because the experiment is carried out on the same equipment, the network delay can be ignored. The system reaction time can be approximately considered as the time required for abnormal detection. Fig. 6 and Fig.7 show the accuracy of detection and the consumption of detection time at this stage respectively.

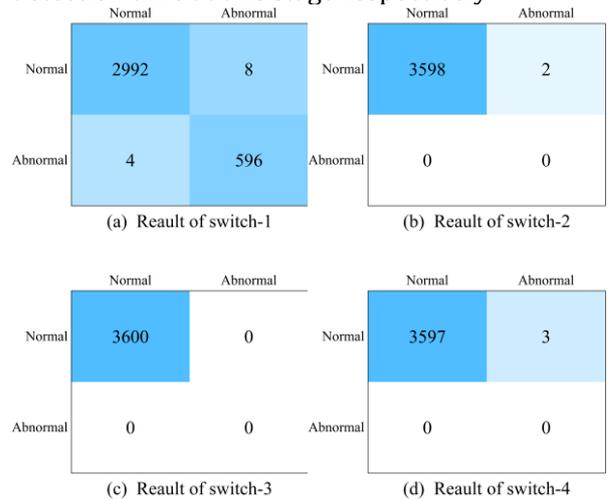

**Fig. 6.** Results of abnormal detection for each switch

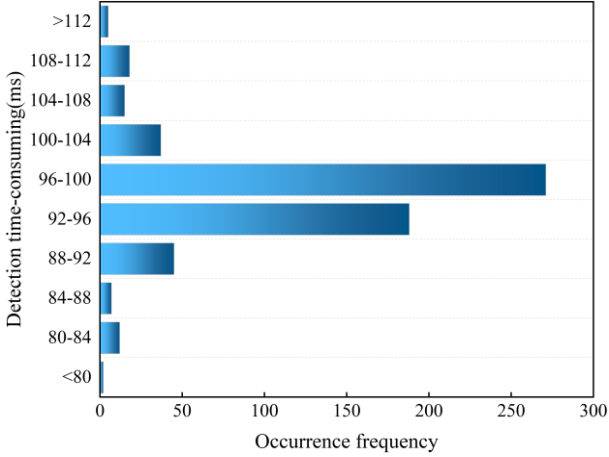

**Fig. 7.** Detection time

As can be seen from Figure 6, in the abnormal detection of port information of four switches, it is found that there are large number of abnormal samples in switch -1, and it can be judged that the switch is an abnormal switch. In addition, the accuracy of detecting abnormal samples in abnormal switches is 0.996 7, the abnormal recall rate is 0.993 3, and the false alarm rate is only 0.002 7. It can be said that the abnormal traffic detection method based on port information can accurately locate the switch connected to the host that launched the DoS attack in SDN and also can accurately judge the attack flow. The reason why Switch-2 and Switch-3 are detected as abnormal is that when normal network services send traffic data with random intensity, the communication traffic suddenly changes temporarily, which is mistaken for abnormal network behavior in the switch control network.

The detection time shown in Fig.7 is mostly within 100ms when an attacker launches a DoS attack, our detection system can find the abnormal behavior of the network and locate the faulty switch within 0.1s This quick response ability can provide support for the real-time security protection research of SDN in the future.

### 5.4.2 Abnormal detection based on traffic data

In order to objectively evaluate the performance of MFDLC, the test experiments at this stage are carried out on InSDN data sets. As can be seen from Table 4, the data set category is unbalanced, and the normal samples and abnormal samples are located in different files. Therefore, we extract 70% of all kinds of samples as the training sets, and the remaining 30% as the test sets. In addition, U2R sample data is too scarce (only 17 samples) to participate in training and testing, so it is excluded. The sample distribution after dataset division is shown in Table 9.

**Table 9** MFDLC hyper-parameter setting

| Sample name | Training set | Test set |
| --- | --- | --- |
| DDoS | 85 359 | 36 583 |
| DoS | 37 531 | 16 085 |
| Probe | 68 690 | 29 439 |
| Brute-Force-Attack | 984 | 422 |
| Web-Attack | 134 | 58 |
| BotNet | 115 | 49 |
| Normal | 47 897 | 20 527 |
| **Total** | **240 710** | **103 163** |

(a) Loss trend of MFDLC

Since no special verification set is divided, the training of MFDLC is carried out by a 10-fold cross-validation method, that is, the training set data is randomly divided into 10 groups, 9 groups are taken out as training data, and 1 group is taken as verification data at a time, and the hyperparameter is determined according to the training results. Then the model is fully trained by using the complete training set data to get the final parameters of the model. Fig. 8 shows the change in the loss function of the MFDLC model during training and testing. When 10 echoes are trained, the training loss decreases rapidly. After training more than 70 echoes, the convergence reaches a stable state, which shows that the model has a stronger ability to resist over-fitting.

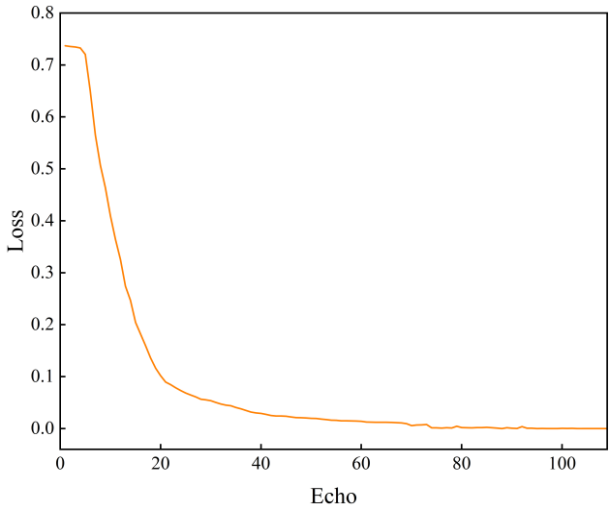

**Fig. 8.** Trend of training loss

(b) Detectability of MFDLC

We preprocess the feature subset of the InSDN test set and input it into the trained MFDLC model for sample classification, and get the classification result as shown in Fig. 9.

**Fig. 9.** Classification performance of MFDLC for InSDN

The highest recall rate of the MFDLC model for each kind of abnormal sample is 99.92%(DDoS) and the lowest is 93.88% (BotNet). However, for the binary classification task that only distinguishes normal samples or abnormal samples, the accuracy rate, precision rate, recall rate, F1 value, and false alarm rate of the model for abnormal detection of the whole test set are 99.83%, 99.93%, 99.86%, and 99.90%, respectively.

(c) Detection performance of different levels of MFDLC

MFDLC is a detection model that adopts three-level wavelet decomposition. To explore the influence of wavelet decomposition on detection performance, we conducted experiments on MFDLC without wavelet decomposition only using the original feature subset and compared the detection results when MFDLC used one-level wavelet decomposition, two-level wavelet decomposition, and four-level wavelet decomposition respectively.

The results are shown in Fig. 10.

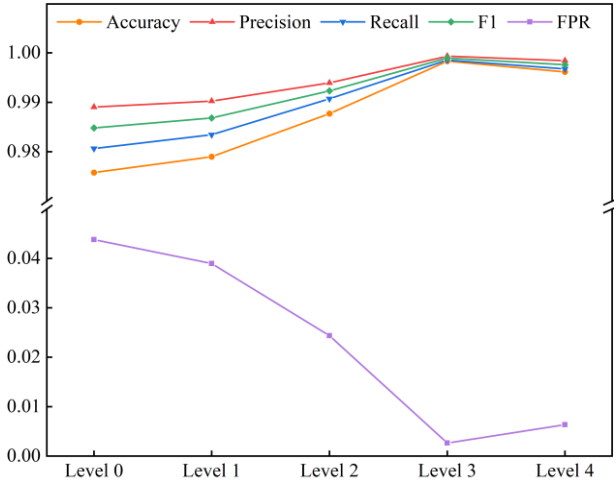

**Fig. 10.** Classification performance of MFDLC under different decomposition scales

With the increase of the decomposition scale, the detection performance of MFDLC is gradually improved. For example, when the decomposition scale is 0, 1, 2, and 3, the accuracy index of MFDLC is 0.9758, 0.9789, 0.9877, and 0.9983 respectively, showing a gradual increase trend. This is because the more scales of wavelet decomposition, the richer information can be provided, and MFDLC can discover more features from sequences with different granularity, which is helpful to improve the model's ability to identify abnormal traffic. However, when the decomposition scale is 4, the accuracy index of MFDLC is 0.9961, which is because the sequence data is redundant after degree decomposition, and then the performance of the classifier is reduced after deep neural network extraction. Increasing the scale of wavelet decomposition continuously can not provide more effective feature information, and the improvement of the detection effect is limited, which is determined by the amount of information contained in the original sample itself.

(d) Compared with other detection methods

In this section, we count some abnormal traffic detection methods in InSDN based on the INSDN data set, including CNN-SoftMax[16] and CNN-LSTM [21]. MFDLC adds multi-scale decomposition and LSTM structure compared with CNN-SoftMax. MFDLC has a similar neural network to CNN-LSTM, but it adds a wavelet decomposition structure. Table 10 clearly shows that MFDLC has higher detection accuracy and lower false alarm rate than traditional detection methods. Compared with the traditional method, the detection accuracy and recall are improved by 1.7% and 1.6%, and the false alarm rate is reduced by 91.3%.

**Table 10** Detection accuracy of different detection methods

| Model | Accuracy | Precision | Recall | F1 | FPR |
|---|---|---|---|---|---|
| CNN-Softmax | 0.985 0 | 0.9827 | 0.9827 | 0.9827 | 0.03 |
| CNN-LSTM | 0.9632 | 0.9760 | 0.9724 | 0.9742 | - |
| MFDLC | 0.9983 | 0.9993 | 0.9986 | 0.9990 | 0.0026 |

(e) Evaluation of generalization ability of MFDLC to different data sets

To accurately test and evaluate the efficiency of the model, we use two data sets commonly used in intrusion detection, namely CIC-IDS2017 and CIC-DoDS2019. They are all feature files extracted by CIC-FlowMeter, so we also select 48 feature subsets of them as our detection data and use the same preprocessing method to prepare the data. We have studied the binary classification performance of two data sets, and the results are shown in Table 11, which clearly shows that our proposed detection model is superior to traditional detection methods on two data sets.

**Table 11** Detection accuracy of different datasets

| Model | DataSet | Acc | Pre | Rec | F1 | FPR |
|---|---|---|---|---|---|---|
| DNN-LSTM | CIC-IDS2017 | 0.993 2 | 0.993 | 0.993 | 0.993 | 0.007 |
| GAN | CIC-DDoS2019 | 0.943 8 | 0.940 8 | 0.978 9 | 0.959 4 | - |
| 1D-CNN& 2D-CNN | CIC-IDS2017 | 0.997 7 | 0.98 | 0.97 | 0.98 | 0.017 4 |
| MFDLC | CIC-IDS2017 | 0.994 5 | 0.997 5 | 0.995 5 | 0.996 6 | 0.009 6 |
| MFDLC | CIC-DDoS2019 | 0.996 1 | 0.997 9 | 0.997 2 | 0.997 6 | 0.008 2 |

In order to verify the generalization ability of the proposed MFDLC to different data sets, we conducted experiments on two public data sets, CIC-IDS2017 and CIC-DDoS2019, and compared the performance with classic detection models such as DNN-LSTM[17], GAN[20] and 1D-CNN&2D-CNN[28]. The accuracy of MFDLC is 0.9945 in CIC-IDS2017, which is 0.13% higher than that of DNN-LSTM, and 5.37% higher than that of GAN in CIC-DDoS2019. Although the accuracy of MFDLC in CIC-IDS2017 is not higher than 0.997 7 of 1D-CNN&2D-CNN, it has achieved higher accuracy and recall rate and has a lower false alarm rate of 0.0096.

## 6  Conclusion

The development of smart cities requires powerful, reliable, and intelligent network services. SDN technology is the trend of future network development, and it needs a fast and efficient means of abnormal detection to maintain its security. Therefore, we propose an effective abnormal traffic detection system, which is able to record only switch port information and form statistics to be submitted to the hierarchical abnormal detection module for the primary detection of abnormal traffic. Experiments show that it can quickly and roughly detect network attacks by calculating the statistical information of switch ports without increasing network components and communication traffic. We also designed MFDLC, which can use wavelet transform and deep learning technology to extract multi-level features from traffic feature data, which is helpful in realizing the SoftMax classifier to classify traffic. The experimental results show that our proposed MFDLC can make full use of the feature information of traffic data, achieve accurate classification of abnormal samples, and achieve higher detection accuracy than traditional detection methods. In addition, MFDLC also shows excellent generalization ability on different datasets. As the scale of smart city networks continues to expand, in the future we will investigate the detection of anomalous traffic in distributed SDN.

## Declaration of competing interest

The authors declare that they have no known competing financial interests or personal relationships that could have appeared to influence the work reported in this paper.

## Data Availability Statement

The data used to support the findings of this study are included within the article.